\documentclass[aps,preprint,onecolumn]{revtex4}
\usepackage{graphicx}
\usepackage{dcolumn}
\usepackage{bm}
\usepackage{amssymb}
\usepackage{amsmath}
\usepackage{amsfonts}

\begin{document}

\title{Concerning an old (but still quite alive) rebuttal of the theorem of John Bell}

\author{Aur\'elien .~Drezet}
\affiliation{Institut N\'{e}el, CNRS and Universit\'{e} Joseph Fourier, BP 166, 38042 Grenoble, France}

\begin{abstract}
In a old  paper by G.~Lochak \cite{Lochak}, it is claimed that the Bell
definition of a hidden variable \cite{Bell1} is in conflict with the
formalism of quantum mechanics. This result implies that it is not
necessary to invoke non locality to explain the violation of the
Bell inequality. A careful analysis of the concept of probability
for hidden variables, as defined differently by Bell and Lochak,
show that
the reasoning and main conclusions of \cite{Lochak} are not correct.\\
Pacs:~03.65.ud
\end{abstract}
\maketitle
\date{\today}

\section{Introduction}
It is well know that de Broglie did not accept the interpretation of
quantum mechanics given by the Copenhagen school of Bohr and
Heisenberg. De Broglie was particulary convinced that a coherent
formulation of quantum mechanics should includes the description of
a dynamical structure at the fundamental level. In this context the
problem of the locality and non locality in quantum mechanics in
connection with the Einstein Podolsky Rosen \cite{EPR} paradox
represented certainly for him a serious challenge. In 1976
G.~Lochak, collaborator of L.~ de Broglie, published in the journal  Foundation of Physics
\cite{Lochak} an argument against the J.~S.~Bell theorem
\cite{Bell1,Bell3,CHSH,Clauser,GHZ,Hardy} concerning local
hidden-variables theories. This refutation was commented briefly by
Bell \cite{Bell2} saying that it may have a mistake in the
reasoning of Lochak since the derivation in \cite{Bell1} identifies
clearly the necessity of non locality in any hidden variables
models. In the words of Bell ``\emph{It may be that Lochak has in
mind some other extension of de Broglie's theory, to the
more-than-one-particle system, than the straightforward
generalization from 3 to 3N that I considered. But if this extension
is local it will not agree with quantum mechanics and if it agrees
with quantum mechanics it will not be local} ''. In the past and
recently Lochak defended his point of view in various articles
\cite{Lochak2} and a Book \cite{Lochak3}. Additionally Lochak
explained to me that de Broglie agreed strongly with his reasoning (this is
confirmed in \cite{Lochak4,Lochak5}). Because of that, and because
of the importance of the de Broglie conceptions in the quest of a
self consistent hidden variable theory, I think that it is necessary
to reexamine the thought of Lochak and de Broglie at that time. I
want to show in this manuscript (written in 2004 but never published) the origin of the problem  by pointing
out different mistakes in the analysis by George Lochak. Beside its historical interest, the (secret) motivation of this work is, as say to me  my mother many times when I was  a child, ``to clean the room before building up something new''.
\section{Refutation of Bell's theorem }
The main idea of \cite{Lochak} is that there is in Bell's reasoning
an hidden statistical assumption which is independent of the
locality condition. Indeed, according to Bell the locality condition
means that there is an hidden variable $\lambda$ used to calculate
all the different measured quantities. Following Bell this hidden
variable must be the initial parameter(s) defining the complete
motion of the system, i.~e.~considered a long time before that the
measurement occurs. Lochak questions this last part of the Bell
definition. Can we really use such parameters when calculating
experimental quantities? Lochak answers by the negative and his
argumentation runs as follow:

Suppose an atomic wave packet propagating along the $z$ axis and
directed on a Stern and Gerlach device such as the one represented
on Fig.~1. A long time before its interaction with the magnetic
field, the initial state is characterized by its wave function that
we suppose to be
$|\psi\rangle=\Psi\left(\mathbf{x},t\right)[c_{\uparrow}|\uparrow\rangle
+c_{\downarrow}|\downarrow\rangle]$. Here we consider only a spin
$1/2$ and $\uparrow,\downarrow$ correspond to the two states of the
spin along $x$.  If the magnetic field is aligned with $x$ the
atomic wave packet is separated into two parts with amplitudes
proportional to $c_{\uparrow}$ and $c_{\downarrow}$ respectively.
Reproducing several times the same experiment shows naturally that
the probability of finding the atom in one of the two exits is
proportional to  $|c_{\uparrow}|^{2}$ and $|c_{\downarrow}|^{2}$
respectively. Similarly changing the direction $\mathbf{a}$ of the
magnetic field implies that the numbers of particles detected in the
two exits become proportional to one or the other of the two
coefficients $|c_{+\mathbf{a}}|^{2}$ and $|c_{-\mathbf{a}}|^{2}$
calculated in the new basis $\pm\mathbf{a}$. Due to the fact that
the spin observable is two valued we will write in the following
$P\left(\alpha=\pm 1,\mathbf{a}\right)$ the single particle
probabilities involved ($\alpha$ are the eigenvalue of the
projection operator $\hat{A}=\boldsymbol{\sigma}\cdot\mathbf{a}$,
where $\sigma_{i}$ ($i=1,2,3$) are Pauli's Matrixes).
\begin{figure}
\includegraphics[width=10 cm]{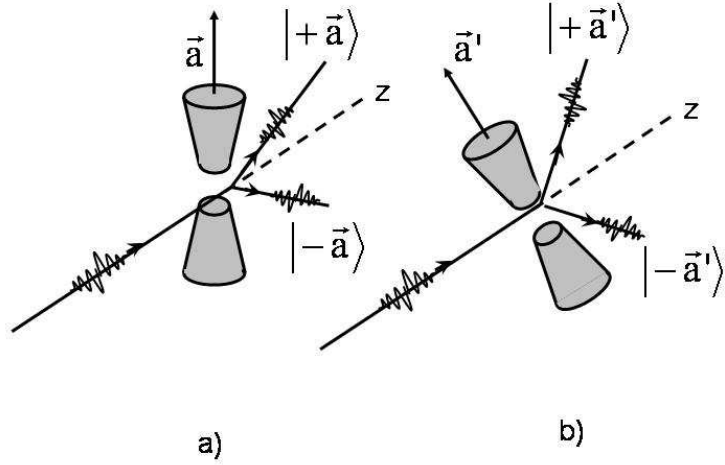}
\caption{A simple idealized apparatus to measure the spin of an atom
of spin $1/2$. Prior to enter in the region of the magnetic field
the atom moves along  the axis z.
After the interaction the  particle is located in one or the other of the two exits corresponding to the
two states of the atomic spin along a) the axis $\mathbf{a}$ or b) the axis $\mathbf{a'}$.}
\end{figure}

Consider now the hidden variable hypothesis as defined by Bell. In
the context of deterministic hidden variable theories, the
probability of finding an atom in one of the two exit doors of a
Stern and Gerlach beam splitter is given by
\begin{eqnarray}
P_{1}\left(\alpha=\pm 1,\mathbf{a}\right)=\int P_{1}\left(\alpha=\pm
1|\mathbf{a},\lambda\right)\rho\left(\lambda\right)d\lambda.\label{e1}
\end{eqnarray}
Here, following Bell, $\lambda $ is associated with the initial
coordinates of the system and we accepted the locality condition
$\rho\left(\lambda,\mathbf{a}\right)\equiv\rho\left(\lambda\right)$
imposing that the initial state is independent of the measurement
settings. G.~Lochak wrote equivalently
\begin{eqnarray}
P_{1}\left(\alpha=\pm 1,\mathbf{a}\right)=\int_{\mathbb{E}_{\alpha,\mathbf{a}}}
\rho\left(\lambda\right)d\lambda\label{e2}
\end{eqnarray}
where $\mathbb{E}_{\alpha,\mathbf{a}}$ is a sub-ensemble of the
hidden variable space associated with the value $\alpha$ taken by
the observable. Such notation is evident if the system follows a
well defined trajectory imposing
$P_{1}\left(\alpha=\pm1|\mathbf{a},\lambda\right)=\delta_{\alpha,A(\lambda,\mathbf{a})}$
 (which by definition of the symbol of Kronecker can only have the
two values 0 or 1 depending on $\lambda$ and $\alpha$). This is
clearly connected to the hypothesis that the observable
$A(\lambda,\mathbf{a})=\pm 1$ must be, as in the de Broglie model,
functions of the initial coordinates. Naturally for another
orientation of the magnetic field we have
\begin{eqnarray}
P_{1}\left(\beta=\pm
1,\mathbf{b}\right)=\int_{\mathbb{E}_{\beta,\mathbf{b}}}
\rho\left(\lambda\right)d\lambda.\label{e3}
\end{eqnarray}
Now comes the paradox enounced by Lochak: Considering the ensembles
intersection
$\mathbb{E}_{\alpha,\mathbf{a}}\cap\mathbb{E}_{\beta,\mathbf{b}}$
allow us to define the probability
\begin{eqnarray}
\mathcal{P}_{1}\left(\alpha
,\mathbf{a},\beta,\mathbf{b}\right)=\int_{\mathbb{E}_{\alpha,\mathbf{a}}\cap\mathbb{E}_{\beta,\mathbf{b}}}
\rho\left(\lambda\right)d\lambda \label{e4}
\end{eqnarray}
i.~e.~
\begin{eqnarray}
 \mathcal{P}_{1}\left(\alpha
,\mathbf{a},\beta,\mathbf{b}\right)=\int\delta_{\alpha,A(\lambda,\mathbf{a})}
\delta_{\beta,A(\lambda,\mathbf{b})}\rho\left(\lambda\right)d\lambda\nonumber\\
=\int
P_{1}\left(\alpha|\mathbf{a},\lambda\right)P_{1}\left(\beta|\mathbf{b},\lambda\right)\rho\left(\lambda\right)d\lambda,
\label{e5}\end{eqnarray}
 which is identified in \cite{Lochak} with the probability of observing
$\alpha$ and $\beta$ with the same particle\cite{remark}. But
quantum mechanics prevents to measure with the same particle values
associated to non-commutative observable. As a consequence, the
assumption that an unique $\rho\left(\lambda\right)$ exists is
erroneous and we must introduce two contextual distributions
$\rho\left(\lambda,\mathbf{a}\right)$ and
$\rho\left(\lambda,\mathbf{b}\right)$. This is in contradiction with
the premisses and we must conclude, if we accept the reasoning of
Lochak, that the definition of Bell \cite{Bell1} is already
unadapted for the description of a single spin measurement. In the
words of Lochak ``\emph{So it was perfectly legitimate and even
evident that the initial probability distribution $\rho(\lambda)$
does not depend on $\mathbf{a}$ (or on $\mathbf{b}$), but this
distribution is not and can not be the one that we need for the
statistics of measurement results: We can be sure of this assertion
precisely because, if we adopt this initial density $\rho(\lambda)$,
we obtain a traditional statistical pattern on measurement results,
which obviously contradicts the well-known and certainly true
statistical results in quantum mechanics}''. The question that we
ask immediately is naturally ``and what about non locality? does it
means that any hidden variable model is necessary non local as seems
to impose to conclude the precedent reasoning? Is a single
particle measurement also non local?''

The answer given by Lochak is that non locality is not necessary
involved for explaining the experimental results. In order to prove that
Lochak analyzed the concrete examples of the L.~de Broglie and
D.~Bohm models \cite{Broglie1,Bohm1,Holland,Bohmfield} which, in the
single particle case, are equivalent and completely local. In these
models a neutral single particle with spin $1/2$, which is
represented by a wave packet having two components
\begin{eqnarray}\Psi\left(\mathbf{x},t\right)=\left(\begin{array}{l}\psi_{\uparrow}\left(\mathbf{x},t\right)\\
\psi_{\downarrow}\left(\mathbf{x},t\right)\end{array}\right),\end{eqnarray}
can be described dynamically as a point like object moving with
the velocity
$\mathbf{v}\left(\mathbf{x},t\right)=[\mathbf{J}/\rho]\left(\mathbf{x},t\right)$.
Here
\begin{eqnarray}
\mathbf{J}\left(\mathbf{x},t\right)=\hbar[|\psi_{\uparrow}\left(\mathbf{x},t\right)|^{2}
\boldsymbol{\nabla}\phi_{\uparrow}\left(\mathbf{x},t\right)+|\psi_{\downarrow}\left(\mathbf{x},t\right)|^{2}
\boldsymbol{\nabla}\phi_{\downarrow}\left(\mathbf{x},t\right)]
\end{eqnarray}
and
\begin{eqnarray}
\rho\left(\mathbf{x},t\right)=|\psi_{\uparrow}\left(\mathbf{x},t\right)|^{2}+|\psi_{\downarrow}\left(\mathbf{x},t\right)|^{2}
\end{eqnarray}
define the probability current and probability density respectively,
and $\phi_{\uparrow},\phi_{\downarrow}$ are the phases of
$\psi_{\uparrow},\psi_{\downarrow}$. In presence of a magnetic field
the two contributions are oriented in one or the other of the exits
\cite{Scully,Holland,Holland2}, separating the trajectories
associated with the two states $\uparrow$ and $\downarrow$.
Naturally, as explained before, any modifications of the magnetic
field orientation change the analyzed basis $\uparrow,\downarrow$.
Consequently in presence of the Stern and Gerlach apparatus
analyzing the spin components along $\mathbf{a}$ and $-\mathbf{a}$
the density of probability
$\rho\left(\mathbf{x},t\right)=|\psi_{\mathbf{a}}\left(\mathbf{x},t\right)|^{2}+|\psi_{-\mathbf{a}}\left(\mathbf{x},t\right)|^{2}$
depends explicitly on the orientation of the magnetic field and must
be written $\rho\left(\mathbf{x},t,\mathbf{a}\right)$. In this model
we can define an instantaneous spin vector
\begin{eqnarray}
\mathbf{S}\left(\mathbf{x},t,\mathbf{a}\right)=
\frac{\Psi^{\dagger}\boldsymbol{\sigma}\Psi}{\rho\left(\mathbf{x},t,\mathbf{a}\right)}.
\end{eqnarray} The projection
$\Sigma\left(\mathbf{x},t,\mathbf{a}\right)=\mathbf{S}\left(\mathbf{x},t,\mathbf{a}\right)\cdot\mathbf{a}$
spans a continuum of values during the interaction with the magnetic
field but at end of the measure (i.~e.~at $t=\infty$) we have
$\Sigma=\pm 1$ corresponding to the spin observable $A=\pm1$. We can
naturally define the mean value of the spin projection $\Sigma$ by
\begin{eqnarray}
E_{1}\left(\sigma\right)=\langle\Psi|\boldsymbol{\sigma}\cdot\mathbf{a}|\Psi\rangle=\int
\Sigma\left(\mathbf{x},t,\mathbf{a}\right)\rho\left(\mathbf{x},t,\mathbf{a}\right)d^{3}\mathbf{x}.
\end{eqnarray}
In the de Broglie theory \cite{Broglie1} the actual position of the
particle is an hidden variable. If we choose to identify the hidden
parameter $\lambda$ used by Bell with the coordinates of the
particle in the wave packet at $t=+\infty$ then we can justify
Eqs.~1,2,3,  with Eq.~10 but in order to do that we must use the
 contextual  distributions $\rho\left(\lambda,\mathbf{a}\right)$ instead
 of the unique $\rho\left(\lambda\right)$ postulated in \cite{Bell1} by Bell. The statistical hypothesis
of Bell is , following \cite{Lochak}, consequently invalidated but
the locality is nevertheless preserved. At that point it is
important to make a break and to consider the problem of the
locality for a single atom in the context of the de Broglie-Bohm
theory. This point is fundamental because Lochak didn't insisted
sufficiently on it in \cite{Lochak}. At the beginning of
\cite{Lochak} we can read that the postulate of locality ``which
will not study further in this paper'' is not a part of the
argumentation. However this postulate is clearly accepted by Lochak
as seen in particular in the sentence ``there is obviously no reason
to suppose any dependance of [the initial] distribution on any
future measurement'', and in ``So it was then perfectly legitimate
and even evident to suppose that the initial probability
dsitribution $\rho(\lambda)$ does not depend on $\mathbf{a}$ (or on
$\mathbf{b}$)'', and again in his discussion with d'Espagnat
\cite{Lochak2}. We accepted as a fact that the model proposed by de
Broglie is local but this need to be verified since in general the
motion of the particle is affected by a quantum potential taking
into account the wave function \cite{Bohm1}. This problem is
discussed in \cite{Holland}. Clearly any modifications of the
boundary conditions or of the magnetic fields during the motion of
the atomic wave packet will disturb the wave function. Such
perturbation propagates into the direction of the particle and
affects the motion later on. There is then effectively no
spontaneous action at distance. In the present case the two spatial
regions occupied by the wave packet and the magnetic field
respectively are well separated at a given time $t=0$ prior to the
measurement. Any instantaneous changes of the orientation of the
Stern and Gerlach device at $t=0$ can generate, in principle, a weak
electromagnetic impulsion propagating in the direction of the atomic
wave packet (a causal signal). The modification of the motion of the
particle by this signal can only occur a time $T= |Z-z(T)|/c$ after
that the change in the measuring device is done ($Z$ is  the
coordinate of the beam splitter on the $z$ axis, $z(T)$ is the
coordinate of the atomic wave packet at $t=T$). This is particulary
clear if we consider that the time evolution of the quantum system
is determined by the interaction hamiltonian which is proportional
to the local value of the magnetic field $\mathbf{B}(\mathbf{x},t)$
at the wave packet position:
\begin{eqnarray}
i\hbar\partial_{t}\psi_{\mathbf{a}}\left(\mathbf{x},t\right)=-\frac{\hbar^{2}\nabla^{2}}{2M}\psi_{\mathbf{a}}\left(\mathbf{x},t\right)
+\mu (\mathbf{B}(\mathbf{x},t)\cdot\mathbf{a})\psi_{\mathbf{a}}\left(\mathbf{x},t\right)\nonumber\\
i\hbar\partial_{t}\psi_{-\mathbf{a}}\left(\mathbf{x},t\right)=
-\frac{\hbar^{2}\nabla^{2}}{2M}\psi_{-\mathbf{a}}\left(\mathbf{x},t\right)
-\mu(\mathbf{B}(\mathbf{x},t)\cdot\mathbf{a})\psi_{-\mathbf{a}}\left(\mathbf{x},t\right)
\end{eqnarray}
($M$ is the atomic mass and $\mu$ the magnetic moment).
 There is then clearly a local dynamic in this problem and the
implicit postulate of Lochak is consequently completely pertinent.

What is the conclusion of Lochak? We saw at a first stage i) that
the definition given by Bell of $\rho(\lambda)$ implies paradoxes
already for the single particle problem. This reasoning shows that
the definition used by Bell imposes the existence of joint
probabilities for non commutating observable in contradiction with
the basic rules of quantum mechanics. At a second stage ii) we saw
that the model of de Broglie, which is local when limited to a
single particle, can explain the result of the single spin
experiment. This proves, if we accept  the reasoning of Lochak, that the density of probability $\rho$ can
depend on $\mathbf{a}$ without introducing nonlocality in the hidden
variable model. Analyzing the de Broglie model Lochak found the
explanation: the variables $\lambda$ are NOT the initial coordinates
of the particle BUT the actual values $\lambda(t)$ of these
coordinates at the time $t$ of the measurement.

Let go back now to the original two particles problem proposed by Bohm.
Two correlated atoms with spin $1/2$ are oriented on two settings of
Stern and Gerlach devices $\mathbf{a},\mathbf{b}$ located apart from
each other. Quantum mechanics allows us to define coincidence
probabilities such as $P_{12}\left(\alpha
,\mathbf{a},\beta,\mathbf{b}\right)$ where $\alpha=\pm1$ are the
eigenvalues of the spin projection operator along $\mathbf{a}$ of
the first atom and similarly $\beta=\pm1$ are the eigenvalues of the
spin projection operator along $\mathbf{b}$ of the second atom.
Using the hidden variables definition given by Bell and applying the
locality condition we write the coincidence probabilities
\begin{eqnarray}
 P_{12}\left(\alpha
,\mathbf{a},\beta,\mathbf{b}\right)=\int
P_{1}\left(\alpha|\mathbf{a},\lambda\right)P_{2}\left(\beta|\mathbf{b},\lambda\right)\rho\left(\lambda\right)d\lambda.\label{e6}
\end{eqnarray}
But now we can repeat the analysis done for the single spin
measurement and define the probabilities
\begin{eqnarray}
 \mathcal{P}_{1}\left(\alpha
,\mathbf{a},\alpha',\mathbf{a'}\right)=\int
P_{1}\left(\alpha|\mathbf{a},\lambda\right)P_{1}\left(\alpha'|\mathbf{a'},\lambda\right)\rho\left(\lambda\right)d\lambda
\end{eqnarray}
and
\begin{eqnarray}
 \mathcal{P}_{2}\left(\beta
,\mathbf{b},\beta',\mathbf{b'}\right)=\int
P_{2}\left(\beta|\mathbf{b},\lambda\right)P_{2}\left(\beta'|\mathbf{b'},\lambda\right)\rho\left(\lambda\right)d\lambda.
\end{eqnarray}
interpreted as the joint probabilities of measuring two non
commutative observable associated with the same particle at the same
time. This is forbidden by quantum mechanics and we must conclude
that $\rho\left(\lambda\right)$ is non adapted to the description of
the experiments. In reality this reasoning is not explicitly present
in \cite{Lochak} but can  be attached to the work of A.~Fine
\cite{Fine}. However Lochak agreed completely with this reasoning
which complete his one\cite{Lochak2}.

Again comes the question: Does this implies necessarily non
locality? Lochak answer a second time by the negative. It could be
that quantum mechanics is effectively non local but it could be that
the hidden variable definition of Bell is simply wrong. Indeed, may be the hidden variable, which are involved in any
calculation of observable and probabilities, should be defined at the
time of the measurements and should include the local settings
$\mathbf{a},\mathbf{b},...$ This direct generalization of the single
particle case to the many particle problem was the aim of the
research made by de Broglie and his group. The fact that de Broglie
never succeeded to build such local model can not however
be considered as a proof against Lochak's reasoning.

\section{Critics and counter arguments}
What I want to show now is that there are several errors in the
deductions of Lochak.

Consider first  the single particle measurement described
in the context of the de Broglie model. Lochak proved that we can
always define the averages and the probabilities as a function of
the actual particle position $\mathbf{x}(t)$ at the time $t$ of the
measurement. If this time tends to $+\infty$ we can justify all the
predictions given by quantum mechanics. However if we consider the
model of de Broglie this is not the unique description of the
phenomenon \cite{Bell3}. Indeed we can always define univocally the
actual position $\mathbf{x}\left(t\right)$ measured for example at
$t=+\infty$ by a function of the initial coordinate
$\mathbf{x}_{0}=\lambda$ of the particle at a time $t_{0}\rightarrow
-\infty$, i.~e.~a long time before that the particle enters in the
Stern and Gerlach apparatus. Due to the conservation of probability requirement
the number of states defined by
$\rho\left(\mathbf{x}_{0},t_{0}\right)\delta^{3}\mathbf{x}_{0}$ in
the elementary volume $\delta^{3}\mathbf{x}_{0}$ is naturally
identical to
$\rho\left(\mathbf{x}\left(t\right),t\right)\delta^{3}\mathbf{x}\left(t\right)$
i.~e.~:
\begin{eqnarray}
\rho\left(\mathbf{x}_{0},t_{0}\right)\delta^{3}\mathbf{x}_{0}
=\rho\left(\mathbf{x}\left(t\right),t\right)\delta^{3}\mathbf{x}\left(t\right).
\end{eqnarray}
This result is of course well known in fluid dynamics where it is associated
to the names of Euler and Lagrange (the so called Euler-Lagrange
coordinates). Similarly
$\Sigma\left(\mathbf{x}\left(t\right),t,\mathbf{a}\right)$ can be
expressed as a function of the initial coordinates of the particle
and can be written
$A\left(\mathbf{x}_{0},t_{0},t,\mathbf{a}\right)$. This is
effectively clear if we write
$\mathbf{x}\left(t\right)=\mathbf{F}_{\mathbf{a}}\left(t,\mathbf{x}_{0},t_{0}\right)$
and substitute it in the expression for $\Sigma$. If we consider now
the expectation value
$\langle\Psi|\boldsymbol{\sigma}\cdot\mathbf{a}|\Psi\rangle$, we can
write
\begin{eqnarray}
E\left(\sigma\right)=\langle\Psi|\boldsymbol{\sigma}\cdot\mathbf{a}|\Psi\rangle=\int
A\left(\mathbf{x}_{0},t_{0},t,\mathbf{a}\right)\rho\left(\mathbf{x}_{0},t_{0}\right)\delta^{3}\mathbf{x}_{0}.
\end{eqnarray}
If we choose $t=+\infty$ then $A=\pm 1$ and we have the complete
definition of Bell (with now $\rho(\lambda)$ independent of
$\mathbf{a}$ as desired). The apparent paradox obtained by Lochak originates from
the fact that we can define the
observables in function  either A) of the initial state or B) of
the actual state obtained after the measurement. This two choices A
and B are mathematically possible and equivalent as it is in classical
fluid mechanics. However, only in the choice A (of Bell) a clear
formulation of locality is possible. Indeed, the Bell definition
of $\rho(\lambda_{0})$ is possible only in a local word but the
definitions $\rho(\lambda, \mathbf{a})$ and $\rho(\lambda,
\mathbf{a},\mathbf{b})$ of Lochak for a system of one or two
particles can be used also in a non local world.
The definitions B of Lochak can therefore not help us to take any
conclusions concerning nonlocality.

There is obviously a contradiction between our conclusion and the
general result of Lochak-Fine concerning the joint probability
$\mathcal{P}_{1}\left(\alpha ,\mathbf{a},\beta,\mathbf{b}\right)$.

In order to solve this paradox it is sufficient to realize that,
contrarily to what it is claimed in \cite{Lochak},
$\mathcal{P}_{1}\left(\alpha ,\mathbf{a},\beta,\mathbf{b}\right)$
has not to be considered as an observable. If we consider the
definition Eq.~5 and interpret this number as a joint probability
associated with an experimental protocol, we deduce of course that
such protocol is realizable if the two measurements are independent
and permutable in opposition with the basics rule of quantum
mechanics. However, the number $\mathcal{P}_{1}\left(\alpha
,\mathbf{a},\beta,\mathbf{b}\right)$ can always be defined
mathematically without any paradox or contradiction and without
being necessarily associated with a possible experiment. In fact,
the ``probability'' $\mathcal{P}_{1}\left(\alpha
,\mathbf{a},\beta,\mathbf{b}\right)$ is just a measure of the number
of common elementary states which go along the direction $\alpha$
when using device $\mathbf{a}$, and along the direction $\beta$ when
using device $\mathbf{b}$, respectively. We can consider the figure
2 as a help: a initial state which is characterized by different
possible $\lambda$ can be represented by an ensemble of point
$\mathbb{E}\{\lambda\}$ characterized by a density $\rho(\lambda)$.
Such ensemble is directly interpreted in the de Broglie model in
which the hidden parameters are the positions of the particles.  For
a given orientation $\mathbf{a}$ of the magnetic field the points of
this ensemble will deterministically move in one or the other of the
separated regions ``+'' and ``-''associated with the two values of
the observable (see figure 2A). By changing the orientation of the
field we modify the partition of the whole ensemble going through
the beam splitter in the region ``+'' or ``-'' (see figure 2B).
However the total number of states is conserved and we can formally
define the number of points
$\int_{\mathbb{E}_{\alpha,\mathbf{a}}\cap\mathbb{E}_{\beta,\mathbf{b}}}
\rho\left(\lambda\right)d\lambda$ contained in the intersections
``$\pm,\pm$'' and ``$\pm,\mp$''associated with these two partitions
of $\mathbb{E}\{\lambda\}$ (see figure 2C). The main thing is that
these new partitions have not to be considered as physical since
they can not be associated with an experimental procedure. This is
the essential mistake of Lochak. More precisely since obviously
$\mathcal{P}_{1}\left(\alpha ,\mathbf{a},\beta,\mathbf{b}\right)$ is
smaller than both $P_{1}\left(\alpha ,\mathbf{a}\right)$ and
$P_{1}\left(\beta ,\mathbf{b}\right)$ \cite{remark3} the number of
points contained in
$\mathbb{E}_{\alpha,\mathbf{a}}\cap\mathbb{E}_{\beta,\mathbf{b}}$
will just contribute partially to the outcomes [$\alpha
,\mathbf{a}$]  and [$\beta ,\mathbf{b}$]. Observing only the total
number of particles contained in these two outcomes (associated with
two  complementary and incompatible experiments) we can consequently
not say if such or such individual particle was or not contained in
$\mathbb{E}_{\alpha,\mathbf{a}}\cap\mathbb{E}_{\beta,\mathbf{b}}$.
In other words this means that the probability
$\mathcal{P}_{1}\left(\alpha ,\mathbf{a},\beta,\mathbf{b}\right)$
defined by Lochak is not an observable but an hidden probability
contrarily to what it was claimed in \cite{Lochak}.
\begin{figure}
\includegraphics[width=10 cm]{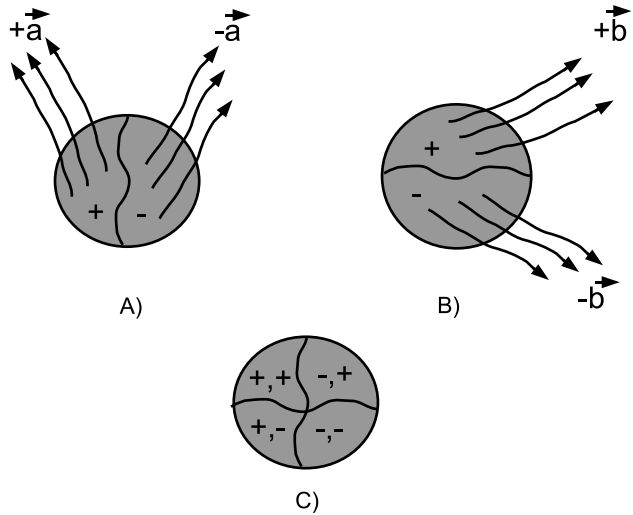}
\caption{Partitions of the whole space $\mathbb{E}\{\lambda\}$ into two parts $+$ and $-$ due to
a measurement of the spin along A)the axis $\mathbf{a}$ or B) the axis $\mathbf{b}$. The intersection
of such two partitions represented on C) is not associated with an individual experience and is just a mathematical definition. }
\end{figure}

It is important to observe that Lochak was very close of the correct
analysis when he said:``\emph{One could object: It is true that we
can not measure in one experiment two different projection of the
spin of [the particle] in the direction $\mathbf{a}$ and
$\mathbf{a}'$, but we can conceive, for the same state $\lambda$ of
this particle, two different experiments for the measurement of each
of these projections and define the probability
$\mathcal{P}\left(\alpha ,\mathbf{a},\alpha',\mathbf{a}'\right)$ of
finding $\alpha$, if we measure the projection $\mathbf{a}$, and
$\alpha'$, if we measure the projection $\mathbf{a}'$}''. However in
spite of his similarity this second interpretation of the
probability proposed in \cite{Lochak} is in fact different of ours
and was rejected by Lochak immediately after its introduction.
Indeed for Lochak `` \emph{two different experiments}'' mean a
temporal succession of two measurements. But it is well know since
the work of de Broglie \cite{Broglie2} that the results of such
succession of experiments depends on the order of the measurement.
We should not then in this context be able to write an unique
expression for the probability $\mathcal{P}_{1}\left(\alpha
,\mathbf{a},\alpha',\mathbf{a}'\right)$ defined in the precedent
citation of Lochak. In the words of Lochak ``\emph{the probability
of finding the value $\alpha$ and $\alpha'$ by measuring two spin
components $\mathbf{a}$ and $\mathbf{a}'$ depends on the order of
the measurements. But this fact and so the nonexistence of the
probability $\mathcal{P}_{1}\left(\alpha
,\mathbf{a},\alpha',\mathbf{a}'\right)$ is not a blemish of the
theory, it is a consequence of the wave particle dualism: If a
hidden-parameter theory contradicts this fact, it will necessary
contradict several correct results of usual wave mechanics}''.
However again there is a misinterpretation and the probability
defined by
$\int_{\mathbb{E}_{\alpha,\mathbf{a}}\cap\mathbb{E}_{\beta,\mathbf{b}}}
\rho\left(\lambda\right)d\lambda$ has not to be associated with a
succession of two measurements. We can define naturally and without
contradiction the probability of two successive measurements of the
spin components using the initial value of the hidden parameter
introduced by Bell and we have
\begin{eqnarray}P_{1}\left(\alpha
,\mathbf{a},\alpha',\mathbf{a}'\right) \int P_{1}\left(\alpha|
\mathbf{a},\alpha',\mathbf{a}',\lambda\right)P_{1}\left(\alpha'|\mathbf{a}',\lambda\right)
\rho\left(\lambda\right)d\lambda.
\end{eqnarray}
Where
$P_{1}\left(\alpha=\pm1|\mathbf{a},\lambda\right)=\delta_{\alpha,A(\lambda,\mathbf{a})}$
is the probability of finding $alpha$ for the first measurement if
we know that the initial value of the hidden parameter is $\lambda$,
and $P_{1}\left(\alpha|
\mathbf{a},\alpha',\mathbf{a}',\lambda\right)$ is the probability of
finding $\alpha'$ for the second measurement if we know that the
initial value of the hidden parameter is $\lambda$ and if we know
that the result of the first measurement was $\alpha$. Clearly we
have no reason to have $P_{1}\left(\alpha|
\mathbf{a},\alpha',\mathbf{a}',\lambda\right)=P_{1}\left(\alpha|
\mathbf{a}\lambda\right)$ because the order of the two measurements
is crucial in the evolution of the dynamical variable. There is then
no reason to identify $P_{1}\left(\alpha
,\mathbf{a},\alpha',\mathbf{a}'\right)$ and
$\mathcal{P}_{1}\left(\alpha
,\mathbf{a},\alpha',\mathbf{a}'\right)$.

It can be added that our reasoning was centered on local
deterministic theories for which we have
$P_{1}\left(\alpha|\lambda,\mathbf{a}\right)=\delta_{\alpha,A(\lambda,\mathbf{a})}$.
However our present refutation can be done in the most general
context of local objective theories \cite{Clauser} for which  the
particle can obey to a statistical hidden dynamic taking into
account the detectors. The probabilities
$P_{1}\left(\alpha|\lambda,\mathbf{a}\right)$ are now general
distributions obeying only to the condition
$P_{1}\left(+|\lambda,\mathbf{a}\right)+P_{1}\left(-|\lambda,\mathbf{a}\right)\leq
1$. We have equality if there is non absorbtion. We see that the
simple interpretation in terms of sub-ensemble given by Lochak in
Eqs.~\ref{e1},\ref{e4} is not possible. However since the second
line of Eq.~\ref{e5} is still valid our critics are the same:
products of probabilities such as
$P_{1}\left(\alpha|\mathbf{a},\lambda\right)P_{1}\left(\beta|\mathbf{b},\lambda\right)$
have not to be interpreted necessary as a probability associated
with a unique experimental process i.~e.~ that we have not
$P_{1}\left(\alpha|\mathbf{a},\lambda\right)P_{1}\left(\beta|\mathbf{b},
\lambda\right)=P_{1}\left(\alpha,\beta|\mathbf{a},\mathbf{b},\lambda\right)$
(which by the way does not exist). A mathematical deduction similar
to the one by Lochak was done by A.~Fine \cite{Fine} and Lochak
refers to his work in his latter publications (See the debate
between d'Espagnat and Lochak on this subject\cite{Lochak2}) but the
conclusion is completely different. Fine concluded that if we use
the definition given by Bell of $\rho(\lambda)$ then we must accept
that we can define probabilities for non commutating observable.
This is identical to the result of Lochak but for Fine this means
that locality must be wrong. However this result is not more
satisfying since it is based on the same misconceptions of the ``
joint '' probabilities involved . Our critics apply then as well to
his reasoning.

Finally I want to make the following remark: i)If we accept that
there is a fundamental dynamic describing the reality (deterministic
or stochastic) in term of a temporal evolution of certain dynamical
parameter, ii) if we accept the principle of locality saying that
the initial state in independent of the subsequent measurement, and
iii) if we accept the basic rules of the probability calculus (in
particular the law of probability conservation), then we can always
write expressions such as \ref{e1} or \ref{e6}. The claim of Lochak
concerning the possibility to save locality by forbidding the use of
the initial density of probability when calculating observable must
then be wrong par principe. Fine didn't make this error but he
identified uncorrectly, as Lochak did before him, the probability
defined by the formula 5,12,13 with a joint probability associated
with an experiment.
\section{Conclusion and remarks}
What is the conclusion of our reasoning? First we observed that the
model of de Broglie considered by Lochak allows the definition of
the expectation values as functions of the initial state
$\lambda_{0}$ existing prior to the measurement. We can in this
description \`{a} la Bell introduce the density of probability
$\rho(\lambda_{0})$ in the calculation of the observable. This is in
contradiction with the claim of Lochak concerning the impossibility
to use such distribution in the calculation of the observable. This
simple fact implies already that the reasoning of Lochak is wrong.
Secondly we found the origin of the paradox observing that the
probabilities defined in Eqs.~4,5 and Eqs.~12,13 have not to be
considered as associated with a measurement procedures. Indeed
quantum mechanics forbids definition of such probabilities in the
case of non commutating observable measurement. The probabilities
Eqs.~4,5,12,13 are in general only mathematical definitions which
have not to be connected with a physical measurement. In other terms
these probabilities are hidden. The problem in the deductions done
by Lochak is even more fundamental and takes his source in a
profound misinterpretation of the definition of a probability space.
Indeed if we accept the concepts of dynamics, locality and of
conservation of probability we must always be able to express any
expectation values as a sum or integral over the initial
distribution $\rho(\lambda_{0})$ of some initial dynamical parameter
$\lambda_{0}$ describing the quantum system. The locality imposes
that the initial state is independent of the observation settings
$\mathbf{a},\mathbf{b}$... and there is consequently no way of
finding a hypothetical loophole in the conclusion given by
Bell\cite{Bell1}.

\section{Historical remark}
Latter after that this manuscript was completed I become aware of a
counter argument of A.~Shimony\cite{Shimony} presenting essentially
the same idea concerning the meaning of the probability
$\mathcal{P}\left(\alpha ,\mathbf{a},\alpha',\mathbf{a}'\right)$.
Detailing his argumentation Shimony explains indeed that well
interpreted $\mathcal{P}\left(\alpha
,\mathbf{a},\alpha',\mathbf{a}'\right)$ is ``\emph{the probability
that the particle is in a state such that a measurement  of spin in
the direction $\mathbf{a}$ (if that option were taken by the
experimenter) would yield the value $\alpha$, and a measurement of
spin in the direction $\mathbf{a}'$ (if that option were taken by
the experimenter) would yield the value $\alpha'$}''.  This
definition is strictly equivalent to ours if we mean by state a
hidden state and if this probability $\mathcal{P}\left(\alpha
,\mathbf{a},\alpha',\mathbf{a}'\right)$ is hidden too. However
Shimony claimed just after that the interpretation that he proposes
was anticipated but rejected immediately by Lochak himself. The
passage of \cite{Lochak} cited by Shimony to prove that is
``\emph{One could object: It is true that we can not measure in one
experiment two different projection of the spin of [the particle] in
the direction $\mathbf{a}$ and $\mathbf{a}'$, but we can conceive,
for the same state $\lambda$ of this particle, two different
experiments for the measurement of each of these projections and
define the probability $\mathcal{P}\left(\alpha
,\mathbf{a},\alpha',\mathbf{a}'\right)$ of finding $\alpha$, if we
measure the projection $\mathbf{a}$, and $\alpha'$, if we measure
the projection $\mathbf{a}'$. This objection is however invalid,
because the impossibility of a simultaneous measurement of two spin
projections of the same particle is not due to a simple
incompatibility of instruments: It comes from the fact that the
state in which we must put the particle to measure its spin
components $\mathbf{a}$ is not the same as the one in which must put
it to measure the component $\mathbf{a}'$}''. Shimony seems not to
have realize that for Lochak the probability
$\mathcal{P}\left(\alpha ,\mathbf{a},\alpha',\mathbf{a}'\right)$ is
a real observable and not a hidden probability . This come from the
fact that the word state used in the Shimony definition can not
refer to a wave function, associated with a experimental procedure,
but only to a pure mathematical subdivision of the whole ensemble
$\mathbb{E}\{\lambda\}$ without direct experimental meaning. For
Lochak, in opposition, this state should be experimentally
accessible and all his critics is erroneously constructed  on this
basis as we explained already before. Here again a not sufficiently
precise definition of the vocabulary used implies some confusions.
It then not surprising that in the rest of his comment Shimony could
not understand Lochak. The last part of the article of Shimony
concerns the example of the Stern and Gerlach measurement described
by Lochak in the context of the double solution theory of de
Broglie. Shimony made here the hypothesis that the difficulty
encountered comes from the fact that the theory of de Broglie takes
into account the disturbance by the measuring device. He guessed
then that the paradox should be solved if instead of considering
the original deterministic hidden variable defined by Bell we used
the most general stochastic theories analyzed in
\cite{Bell3,Clauser}. This is unfortunately wrong because the de
Broglie model is strictly deterministic.


\begin{thebibliography}{}
\bibitem{Lochak}
G.~Lochak, Found.~Phys.~\textbf{6},173 (1976). This article was
published in french in Epistemological Letters (Bienne) \textbf{38},
41 (1975).
\bibitem{Bell1}
J.~S.~Bell, Physics~\textbf{1},195 (1964).
\bibitem{EPR}
A.~Einstein, B.~Podolsky and N.~Rosen, Phys.~Rev.~\textbf{46}, 777
(1935).
\bibitem{Bell3}
J.~S.~Bell, \emph{Introduction to the hidden variable question, in:
Proceedings of the international school of physics Enrico Fermi,
course IL}, Academic Press Inc.~, New York, 1971.
\bibitem{CHSH}
J.~Clauser, M.~A.~Horne, A.~Shimony and R.~A.~Holt,
Phys.~Rev.~Lett.~\textbf{23}, 880 (1969).
\bibitem{Clauser}
a)J.~F.~Clauser, M.~A.~Horne, Phys.~Rev.~\textbf{D10}, 526 (1974).\\
b)J.~F.~Clauser, A.~Shimony, Rep.~Prog.~Phys.~\textbf{41}, 1881
(1978).
\bibitem{GHZ}
D.~M.~Greenberger, M.~A.~Horne, A.~Shimony, and A.~Zeilinger,
Am.~J.~Phys.~\textbf{58}, 1131 (1990).
\bibitem{Hardy}
L.~Hardy, Phys.~Rev.~Lett.~\textbf{72}, 781 (1994).
\bibitem{Bell2}
J.~S.~Bell, \emph{Speakable and unspeakable in quantum mechanics},
Cambridge University Press, Cambridge, 1987, pp.~63-66. This article
reproduces the answer given by Bell to Lochak which was originally
published in Epistemological letters (Bienne) \textbf{7}, 2 (1976).
\bibitem{Lochak2}
 See G.~Lochak, Ann.~Fond.~Louis de Broglie \textbf{26}, 5 and 31 (2001)
 for a complete list of
 references, and in particular G.~Lochak Revue de m\'etaphysique et de
 morale 1/1983 p.~85 and 1/1985 p.~400.
\bibitem{Lochak3}
G.~Lochak, \emph{Louis de Broglie}, Champs-Flammarion , Paris,
1992, pp.~248-249.
\bibitem{Lochak4}
L.~de Broglie, G.~Lochak, J.~A.~Beswick, and
 J.~Vassalo-Pereira, Found.~Phys.~\textbf{6}, 3 (1976).
\bibitem{Lochak5}
G.~Lochak private communications.
\bibitem{remark}
Other equivalent definitions are
\begin{eqnarray}
P(+,\mathbf{a},+,\mathbf{b})=\int d\lambda\rho\left(\lambda\right)
\frac{\left(A\left(\lambda,\mathbf{a}\right)+1\right)}{2}\frac{\left(A\left(\lambda,\mathbf{b}\right)+1\right)}{2}\nonumber\\
P(+,\mathbf{a},-,\mathbf{b})=\int d\lambda\rho\left(\lambda\right)
\frac{\left(A\left(\lambda,\mathbf{a}\right)+1\right)}{2}\frac{\left(1-A\left(\lambda,\mathbf{b}\right)\right)}{2}\nonumber\\
P(-,\mathbf{a},+,\mathbf{b})=\int d\lambda\rho\left(\lambda\right)
\frac{\left(1-A\left(\lambda,\mathbf{a}\right)\right)}{2}\frac{\left(A\left(\lambda,\mathbf{b}\right)+1\right)}{2}\nonumber\\
P(-,\mathbf{a},-,\mathbf{b})=\int d\lambda\rho\left(\lambda\right)
\frac{\left(1-A\left(\lambda,\mathbf{a}\right)\right)}{2}\frac{\left(1-A\left(\lambda,\mathbf{b}\right)\right)}{2},\nonumber
\end{eqnarray} where $\pm$ refer to the two values $\pm 1$ of the
observable $A\left(\lambda,\mathbf{a}\right)$ and
$A\left(\lambda,\mathbf{b}\right)$.
\bibitem{Broglie1}
a)L.~de Broglie, C.~R.~Acad.~Sci.~Paris 183 (1926) 447; 185 (1927)
580.\\b)L.~de Broglie, Nonlinear wave mechanics, Elsevier,
Amsterdam, 1960.
\bibitem{Bohm1}
D.~Bohm ,  Phys.~Rev.~\textbf{85}, 166 and 180 (1952).
\bibitem{Holland}
P.~R.~Holland, The Quantum Theory of Motion, Cambridge University
Press, Cambridge, 1993.
\bibitem{Bohmfield}
D.~Bohm, B.~J.~Hiley and P.~N.~Kaloyerou, Phys.~Rep.~144 (1987) 321.
\bibitem{Scully}
M.~O.~Scully, W.~E.~Lamb,
and A.~O.~Barut, Found.~Phys.~\textbf{17}, 575 (1987).
\bibitem{Holland2}
C.~Dewdney, P.~R.~Holland and A.~Kyprianidis
Phys.~Lett.~\textbf{A119}, 259 (1986).
\bibitem{Fine}
A.~Fine, Phys.~Rev.~Lett.~\textbf{48}, 291 (1982).
\bibitem{remark3}
We have $\mathcal{P}_{1}\left(\alpha ,\mathbf{a},\beta,\mathbf{b}\right)+\mathcal{P}_{1}\left(\alpha ,\mathbf{a},-\beta,\mathbf{b}\right)=P_{1}\left(\alpha,\mathbf{a}\right)$ and similarly $\mathcal{P}_{1}\left(\alpha,\mathbf{a},\beta,\mathbf{b}\right)+\mathcal{P}_{1}\left(-\alpha,\mathbf{a},\beta,\mathbf{b}\right)=P_{1}\left(\beta,\mathbf{b}\right)$.
\bibitem{Broglie2}
L.~de Broglie,  La th\'{e}orie de la mesure en m\'{e}canique
ondulatoire (interpr\'{e}tation usuelle et interpr\'{e}tation
causale), Gauthier-Villars, Paris 1957.
\bibitem{Shimony}
A.~Shimony, Epistemological Letters \textbf{8}, 1 (1976).

\end{thebibliography}
\end{document}